\documentclass{PoS_arXiv}

\usepackage{cite}
\usepackage{graphicx}
\usepackage{amssymb}
\usepackage{amsmath}
\usepackage{bbm}
\usepackage{cancel}
\usepackage{url}
\usepackage{bbold}
\usepackage{stmaryrd}
\usepackage{ulem}

\def\be{\begin{equation}}
\def\ee{\end{equation}}
\newcommand{\ol}{\overline}

\newcommand{\bea}{\begin{eqnarray}}
\newcommand{\eea}{\end{eqnarray}}
\newcommand{\amc}{{\sc MadGraph5\_aMC@NLO}}
\newcommand{\fr}{{\sc FeynRules}}
\newcommand{\ufo}{{\sc UFO}}
\newcommand{\nlo}{{\sc NLOCT}}
\newcommand{\fa}{{\sc FeynArts}}
\newcommand{\ml}{{\sc MadLoop}}
\newcommand{\alo}{{\sc ALOHA}}

\title{Automated Two Higgs Doublet Model at NLO}

\ShortTitle{Automated Two Higgs Doublet Model at NLO}

\author{\speaker{Celine DEGRANDE}\\%
        IPPP - University of Durham\\
        E-mail: \email{celine.degrande@durham.ac.uk}}

\abstract{The Two Higgs Doublet Model at NLO is generated automatically by \fr\ and \nlo\ and allows any computation to be performed at NLO in QCD inside \amc. The model can handle both four and five massless flavours. Preliminary results of the shape comparison between the two schemes are shown.}

\FullConference{Prospects for Charged Higgs Discovery at Colliders\\
                 16-18 September 2014\\
                 Uppsala University Sweden                 }

\begin{document}

\section{Introduction}

Charged Higgs production at the LHC like many other processes receives large corrections from the strong interaction. Consequently, the Next to Leading Order (NLO) predictions are desirable for the searches for a charged scalar. So far charged Higgs production in the context of the Two Higgs Doublet Model (2HDM) or the Minimal Supersymmetric Standard Model (MSSM) has only been implemented by hand~\cite{Weydert:2009vr,Beenakker:1996ed,Prospino2}.
On the contrary, any Standard Model (SM) process can now be computed at the NLO in QCD using \amc~\cite{Alwall:2014hca}. This situation is not owned by charged Higgs production only but is common for many processes in extensions of the SM.
This issue is purely related to the loop amplitudes as any computation requiring the tree-level amplitudes can be performed automatically once  the BSM Lagrangian has been introduced in \fr~\cite{Alloul:2013bka} for example. As a matter of fact, all the tree-level vertices are then extracted and sent in the appropriate format to one of the various tools handling the computation of the tree-level matrix element. 
However, two additional ingredients are needed for the computation of the one-loop amplitudes by \ml~\cite{Hirschi:2011pa}  in \amc. The first one is the UV counterterms, i.e. the vertices introduced by the renormalisation of the Lagrangian to cancel all the UV divergences appearing in the loop. The second missing element is the so called rational $R_2$ vertices. 
Any one-loop amplitude can be written as 
\be
A  = \sum_i d_i \text{Box}_i + \sum_i c_i \text{Triangle}_i + \sum_i b_i \text{Bubble}_i + \sum_i a_i \text{Tadpole}_i +R,
\ee 
where the Box, Triangle, Bubble and Tadpole are known scalar integrals in four dimension and $R$ is the rational term. The rational terms are finite contributions generated by the $d-4$ dimensional part of the integrand. They are split in two parts, the $R_1$ originating from the denominators of the integrand and the $R_2$ generated by its numerator. While the $R_1$ terms can be computed similarly as the four dimensional part of the amplitude but using a different set of known integrals, the $R_2$ terms require an analytical computation. The $R_2$ terms for any amplitude can nevertheless be computed from a set of process independent vertices that should be provided in the \ufo\ model~\cite{Degrande:2011ua} to \amc.

\section{Automatic generation of \ufo\ at NLO}

The aim of the recent version of \fr\ and the new \nlo\ package is to compute the missing ingredients for the evaluation of the loop amplitude  and to include them in the UFO model automatically~\cite{Degrande:2014vpa}. The renormalization is performed inside \fr\ where the Lagrangian as well as all the relations between the parameters are known by the {\tt OnShellRenormalization} function:
\begin{verbatim}
Lren = OnShellRenormalization[MyLag, options]
\end{verbatim}
where {\tt MyLag} is the Lagrangian of the model in \fr\ notation. The renormalized Lagrangian is then passed to \fa~\cite{Hahn:2000kx} using the corresponding \fr\ interface. The \nlo\ package uses \fa\ to write the irreducible one-loop amplitudes and then computes the $R_2$ vertices and the UV counterterms by solving the renormalization conditions. The evaluation of all the vertices is performed by a single function
\begin{verbatim}
WriteCT[<model>,<genericfile>,options]
\end{verbatim}
where {\tt model} and {\tt genericfile} are the \fa\ input files. Additionally, 
the resulting NLO vertices are written in a text file (with a {\tt nlo} extension). The options of those two functions are detailed in Ref.~\cite{Degrande:2014vpa}. This file is finally loaded into \fr\ after the model such that the \ufo\ with the one-loop ingredients can be written by
\begin{verbatim}
WriteUFO[MyLag, UVCounterterms -> UV$vertlist, 
R2Vertices -> R2$vertlist].
\end{verbatim}
where \verb+ UV$vertlist+ and \verb+ R2$vertlist+ contain respectively the UV counterterm vertices and the $R_2$ vertices and are both defined in the  {\tt .nlo} file. \\
The only model restriction is that the Lagrangian should be renormalizable, i.e. there should not be any operator of dimension higher than 4. The model should also be implemented in the Feynman gauge for the \nlo\ package  as well as \amc\ to work properly.
The masses and wave functions are chosen to be renormalized in the on-shell scheme while all the other external parameters are renormalized in the $\ol{MS}$. However, the zero momentum scheme can be chosen for the gauge coupling of a massless gauge boson. \\
The method has been validated by comparing analytically the $R_2$ vertices of the SM generated by both the strong and the electroweak interactions and of the MSSM from the QCD one-loop corrections to the expressions in the literature~\cite{Draggiotis:2009yb,Garzelli:2009is,Shao:2012ja}. The UV counterterm vertices have also been compared analytically to the expressions in ~\cite{Beenakker:2002nc,Denner:1991kt} for all the SM one-loop contributions.  
The results of many processes for SM \ufo\ with NLO vertices from the strong interaction has also been found in agreement with the built-in version of of the SM in \amc.
Finally, pole cancelation between the real and the virtual and gauge invariance\footnote{Only the sum of the $R_1$ and $R_2$ terms is a gauge invariant quantity.} have been used to check the 2HDM as well as other BSM models.

\section{The 2HDM at NLO}

\begin{figure}[t]
\centering
\includegraphics[width=0.7\textwidth]{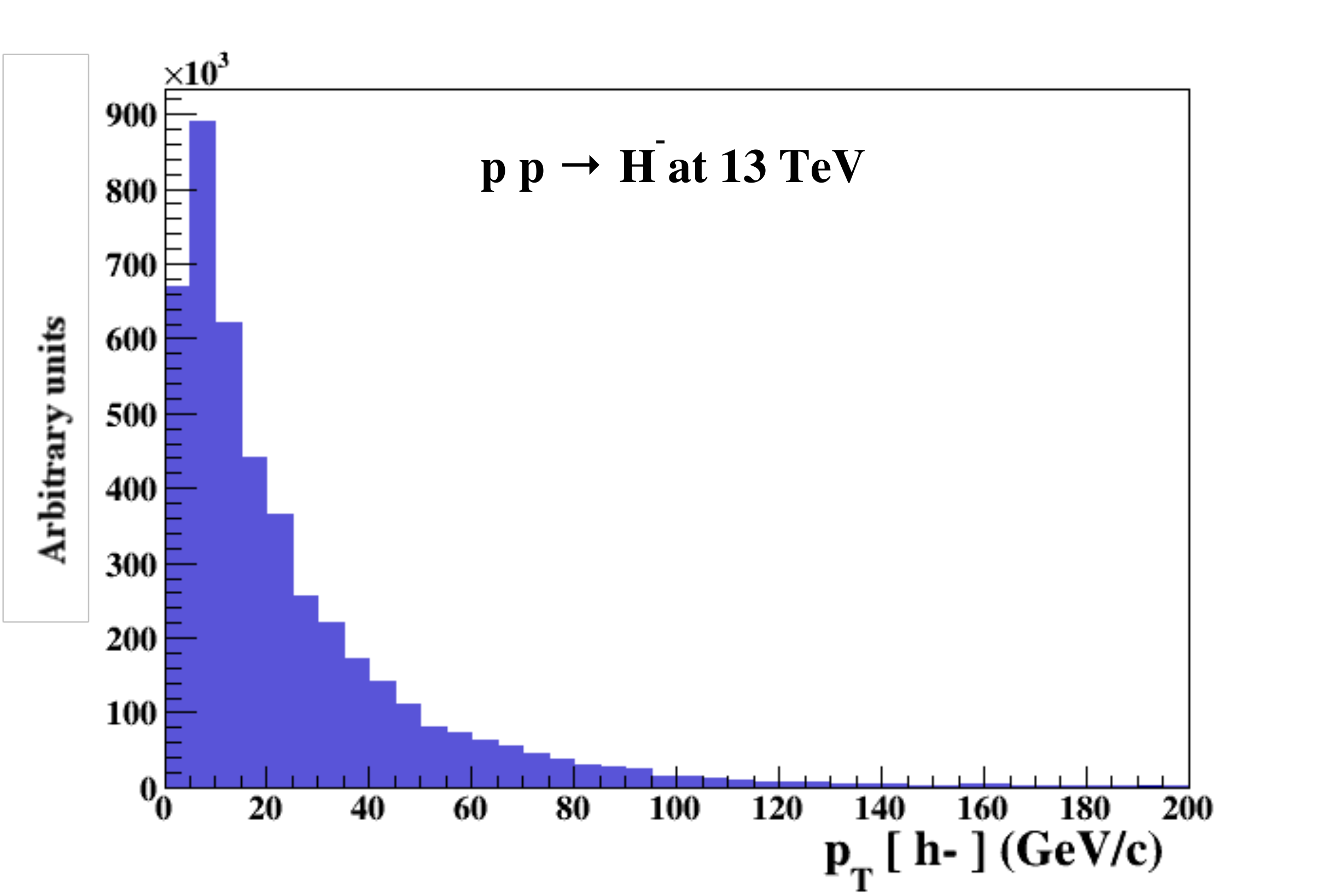}
\caption{$p_T$ distribution of the charged Higgs from flavour changing interactions between the first two generations of quarks: $G^U_{12}=G^U_{21}=G^D_{12}=G^D_{21}$}
\label{fig:FChm}
\end{figure}

\begin{figure}[!ht]
\centering
\includegraphics[trim=0 300 0 0, clip,width=0.7\textwidth]{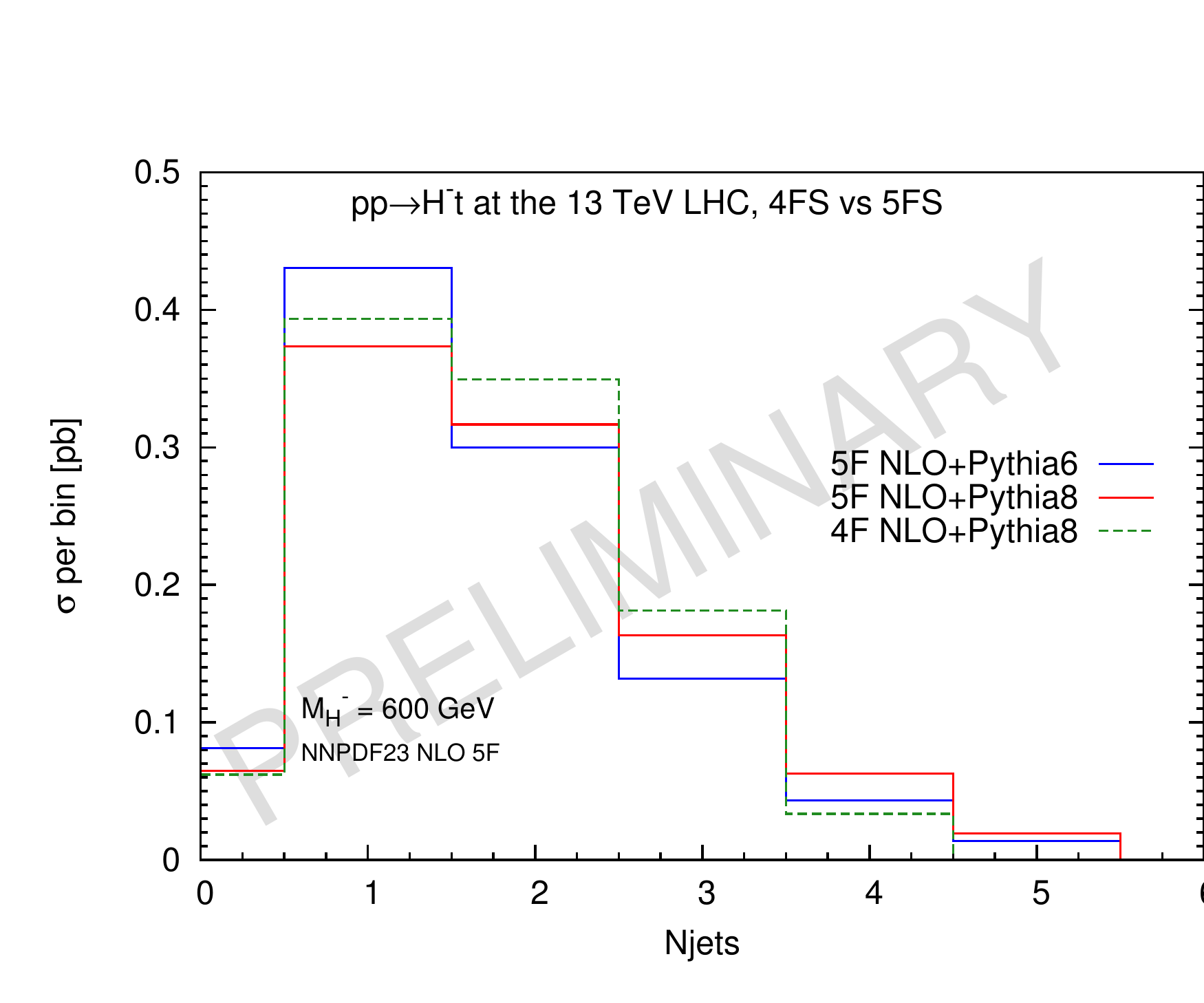}
\caption{Number of jets distribution for the charged Higgs production in association with a top quark in the five and four flavours schemes.}
\label{fig:thm}
\end{figure}

A new implementation of the generic 2HDM into \fr\ has been done such that all the relation between the parameters are present in the model file as required for the renormalization of the Lagrangian. Namely, the number of independent parameters of the model is equal to the number of external parameters. Therefore, no external tool is needed to compute the values of the parameters appearing in the input files.  To ease the on-shell renormalization, all the masses of the physical scalars have been implemented as external parameters. Since the model is implemented in the Higgs basis, the couplings of the doublet with a vacuum expectation value (v.e.v.) to the fermions is directly related to their masses. On the contrary, the fermion couplings to the other scalar doublet are free parameters. Therefore the renormalisation of the first ones is fixed by the renormalization of the fermion masses and is done in the on-shell scheme while the latter are renormalized in the $\ol{MS}$ scheme. Consequently, the model can only be transformed to a type I or a type II 2HDM valid at NLO at the \fr\ level and before the renormalization.\\
Although the \ufo\ model can be generated with one-loop ingredients for both the electroweak and the strong corrections, only the corrections from the strong interaction can be handled by \amc\ so far.  Nevertheless, any process in the 2HDM can be studied now at the NLO in QCD thanks to  the full \fr-\nlo- \fa-\ufo-\alo~\cite{deAquino:2011ub}-\amc\ chain. \\
The simplest way of producing a charged higgs is by the annihilation of an up-type quark and down-type quark. For example, the Higgs $p_T$ distribution for $p p \to H^-$ is displayed in Fig.~\ref{fig:FChm}. The charged Higgs is produced by allowing yukawa interactions mixing the first two generations of quarks. Despite that those couplings are free parameters in the generic 2HDM, they are strongly constrained by the experimental data on flavour changing neutral currents. 
Those constraints are avoided theoretically by allowing each fermion type to couple only one scalar doublet like in the type I or II 2HDM. The yukawa couplings are then all related to the masses and the dominant production mechanism for a heavy charged Higgs is the associated top production.
The 2HDM of type II can then be used to study the shape difference between the five and four flavours schemes in the production of a charged Higgs in association with a top quark~\cite{4and5F} in the same spirit as what has been done for the total cross-section~\cite{Flechl:2014wfa}.  The preliminary distribution for the number of jets is shown in Fig.~\ref{fig:thm} as an illustration. While by default the bottom Yukawa is renormalized in the on-shell scheme, the $\ol{MS}$ is more suitable to avoid large logarithm. 

\section{Summary}

The 2HDM \ufo\ model with all the ingredients required by \amc\ to generate any process at the NLO in QCD has been produced automatically by \fr\ and \nlo. In particular, this model can be used to study charged Higgs production at NLO both in the four and five flavours schemes. The method is fully automated and can therefore easily be used to study the production of charged scalar particles in other extension of the SM as well.

\acknowledgments
The author is a Durham International Junior Research Fellow.

\bibliographystyle{JHEP}
\bibliography{biblio}

\end{document}